\documentclass[12pt]{iopart}

\usepackage{graphicx}

\begin{document}
\pagenumbering{arabic}

\title[]{Density functional theory study of quasi-free-standing graphene layer on 4H-SiC($0001$) surface decoupled by hydrogen atoms}
\author{Jakub So\l{}tys $^1$, Jacek Piechota $^1$ , Micha\l{} \L{}opuszy\'nski $^1$  and Stanis\l{}aw Krukowski $^{1,2}$ }
\address{%
$^1$ Interdisciplinary Center for Materials Modeling, University of Warsaw, Pawi\'nskiego 5a, 02-106 Warsaw, Poland  \\
$^2$ Institute of High Pressure Physics, Polish Academy of Sciences, Soko\l{}owska 29/37, 01-142 Warsaw, Poland 
}%

\begin{abstract}
Epitaxial graphene, grown on SiC($0001$) surface, has been widely studied both experimentally and theoretically.
It was found that first epitaxial graphene layer in such structures is a buffer layer i.e. there are no characteristic
Dirac cones in the band structure associated with it.  However,  C. Riedl et al. (Phys. Rev. Lett. 103, 246804
(2009)) in their experimental work observed recently that hydrogen intercalation of SiC-graphene samples can
recover electronic properties typical to selfstanding graphene. The possible scenarios of hydrogen intercalation inducing graphene layer decoupling, including both the hydrogen penetration paths and energetically stable positions of hydrogen atoms, were modeled in \textit{ab initio} DFT calculations. 
From the obtained results it follows that, due to intercalation, the graphene layer moves away to achieve about 3.9 \AA \ distance from the SiC surface. 
Electronic band structure, calculated for such quasi free standing graphene, exhibits Dirac-cone behavior
which is in agreement with ARPES measurements.  
\end{abstract}

\maketitle

Graphene is the one of the most interesting topic in recent decades. 
It is because of its outstanding electronic properties such as massless Dirac fermions, 
large electron coherence lengths,  anomalous integer Hall effect \cite{Berger,Novoselov,Zhang},
ballistic transport at room temperature, and good capability of integration with the silicon planar technology. 
Therefore, graphene is also promising material for wide range of applications, such as high speed electronics, detection of some species, etc. \cite{Geim, Berger}. Selfstanding graphene is not stable mechanically, therefore very promising route was found when it was discovered that growth of graphene on SiC substrate is relatively simple. 
It was shown that graphene growth on SiC surface is possible on both Si and C-faces. 
High quality graphene layers on SiC($0001$), i.e on Si-face side are smooth and homogeneous \cite{Emtsev2}.
Possibility of fabrication of large area epitaxial graphene and its integration with existing device
technologies drawn attention of many researchers \cite{Berger, Ohta, Riedl2} .	
However, the first graphene layer formed on SiC($0001$) surface consists of $sp^2$ and 
$sp^3$ hybridized carbon atoms \cite{Emtsev3, Varchon},
therefore first graphene layer is covalently bound to SiC($0001$) surface.
This drastically lowers the mobility of carriers which does not exceed 2.000 
$\rm{cm}^{-2} \rm{V}^{-1} \rm{s}^{-1}$ at low temperatures \cite{Emtsev2}.
Experimental \cite{Borysiuk} and theoretical \cite{Mattaush} studies are consistent. They show 
that graphene-SiC($0001$) surface distance is about 2 \AA. 
\textit{Ab initio} studies indicate that the first carbon layer is a buffer layer and there are 
no Dirac cones in its band structure. This explains low carrier mobility, measured experimentally.
However, recent experimental work show that  electronic properties typical for graphene can be recovered after 
hydrogen intercalation of graphene single layer, deposited on SiC($0001$) surface. Riedl et. al. in their work \cite{Riedl} prepared epitaxial graphene on 4H-SiC($0001$) surface. From these samples, they were able to obtain quasi-free-standing epitaxial graphene after annealing in molecular hydrogen at atmospheric pressure at the temperature between $600^{\circ}$C and $1000^{\circ}$C. 
Inspired by their results we have modeled possible hydrogen intercalation scenarios
and penetration paths of hydrogen atoms.

\begin{figure}
\begin{center}
\includegraphics[width=0.7 \textwidth]{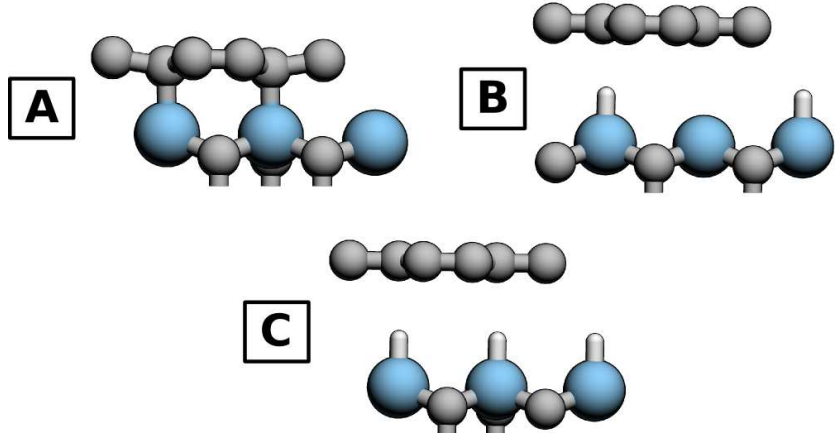}%
\end{center}
\caption{\label{fig:hydro} (Color online) Graphene-SiC interface before and after hydrogen treatment \cite{qutemol}. 
When hydrogen atoms are absent graphene in covalently bound to the surface. Two graphene atoms
which are involved in binding are shifted towards the surface. Bond length between graphene 
layer and the surface is 2.0 \AA \ (A) . 
After hydrogen intercalation with two (B) or three (C) atoms, covalent bonds are broken and
graphene layer is decoupled from the surface. Graphene surface is flat and about 3.70 \AA \ (B) , 3.9 \AA \ (C) , 
above the SiC surface  }
\end{figure}

In this Letter we employ \textit{ab initio} density functional theory to investigate
graphene-SiC interface. 
In all reported calculations  VASP \cite{Kresse1993, Kresse1996, Kresse1996a, Kresse1999} code was employed.  The projector augmented wave (PAW) approach \cite{Blochl}
was used in its variant available in the VASP package \cite{Kresse1999}.
For the exchange-correlation functional local spin density approximation (LSDA) 
was applied.
A plane wave cutoff energy was set to 500 eV. Monkhorst-Pack k-point mesh was set to $7\times7\times1$. 
4H-SiC(0001) superlattice was constructed using 8 bilayers of Si-C, which sufficiently well
approximates the properties of real surface.  
At the top of the SiC structure one carbon atomic layer was placed. 
The slab separation space width was varied, depending on the considered case (with or without hydrogen), between
26.8 and  28.5 \AA \ .
Due to the lattice mismatch between SiC and graphite we have performed elastic adjustment at the interface. 
Two top SiC layers and graphene layer was relaxed. Conjugate gradient algorithm was used in the relaxation of the atomic positions.

We have used the model proposed by Mattausch and Pankratov e.g.  $ \sqrt{3} \times \sqrt{3} R 30 ^\circ - SiC$ 
unit cell with fitted graphene layer (GL). In this case GL is covalently bonded to the substrate. 
Covalent bonds overcompensates the elastic stress at the interface. 
Previous works showed that covalent bonding of the substrate removes the graphene-type electronic 
features from the energy region around the Fermi level. After relaxation the bond length between SiC surface and GL was about 2 \AA \ which is in agreement with earlier theoretical and experimental works. 
In this model only two of the three topmost Si atoms are covalently bound to the graphene layer,
one Si atom remains unbounded.

Subsequently the system with the hydrogen molecule positioned under the graphene layer, was simulated.
It turned out that relaxation pattern is as follows: 
one hydrogen  atom from H$_2$ molecule is detached and subsequently it is attached to the free-standing top Si atom. The remaining one breaks the SiC-graphene bond. In the result, the graphene layer is decoupled from the surface, it moves away to the distance of 3.7 \AA \ from the SiC surface (see. Fig. \ref{fig:hydro} B) . 
The graphene layer in this configuration in not covalently bound to the surface but 
it is quasi free standing over SiC surface.
Fig. \ref{fig:hydro} C presents the situation in which the GL and SiC surface is separated by 
hydrogen atoms which are breaking covalent Si-graphene bonds. The most stable position of 
hydrogen atoms is perpendicular to the SiC surface.  
In this model graphene layer is about 3.9 \AA \ above the surface.
Despite the distance, the graphene monolayer still weakly interacts with the surface. 
Riedl et al. \cite{Riedl} observed that LEED patterns were suppressed after hydrogen treatment 
which indicates much smaller displacements in graphene layer. 
Comparing situations presented in Fig. \ref{fig:hydro} A, B and C we note that in B and C case graphene
layer is almost flat, which is in contrast with the A case where the covalently bonded atoms are shifted towards the surface.

\begin{figure}[b]
\begin{center}
\includegraphics[width=0.7 \textwidth]{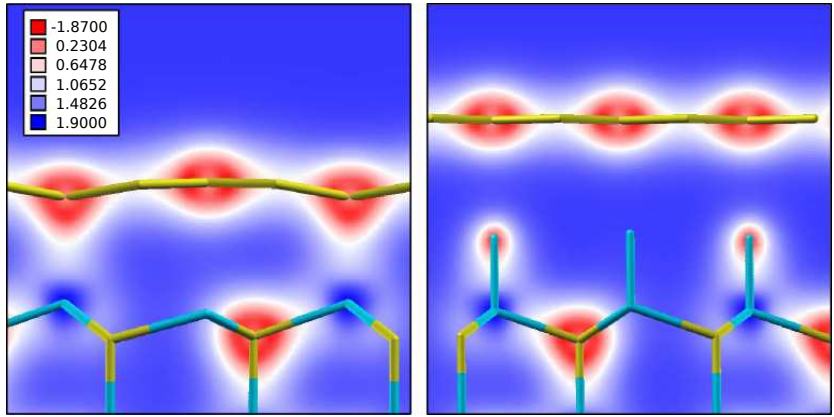}%
\end{center}
\caption{\label{fig:chgcar} (Color online) Charge density [el/\AA$^{-3}$] cross-section (passing through bound graphen atoms )
in the graphene-SiC interface before (left) and after (right) hydrogen treatment \cite{xcrysden}. 
Two carbon atoms of the graphene layer are covalently bound to the surface; they are moved closer to the surface.
Electron density around these atoms is strongly affected by the bonding to the Si surface atoms. The layer is now formed by $sp^2$ (not bound) and $sp^3$ (bound) hybridized 
carbon atoms. After hydrogen treatment charge density distribution is quite different,
graphene layer is flat and the charge density for each atom in the layer is identical. Graphene electronic properties (Dirac cones) are now recovered.
  }
\end{figure}

The analysis of electron density profiles can give further insight into nature of intercalation process.
Charge density maps (see. Fig. \ref{fig:chgcar}) show that in the case of graphene covalently bound to 
the surface graphene atoms are strongly affected by the SiC substrate and are now in two configurations 
- $sp^2$ (not bound to the surface) and $sp^3$ (bound to the surface). Situation is quite different after 
hydrogen treatment.  In this case each graphene atom, as in isolated graphene layer, has $sp^2$ configuration. 

\begin{figure}
\begin{center}
\includegraphics[width=0.6 \textwidth]{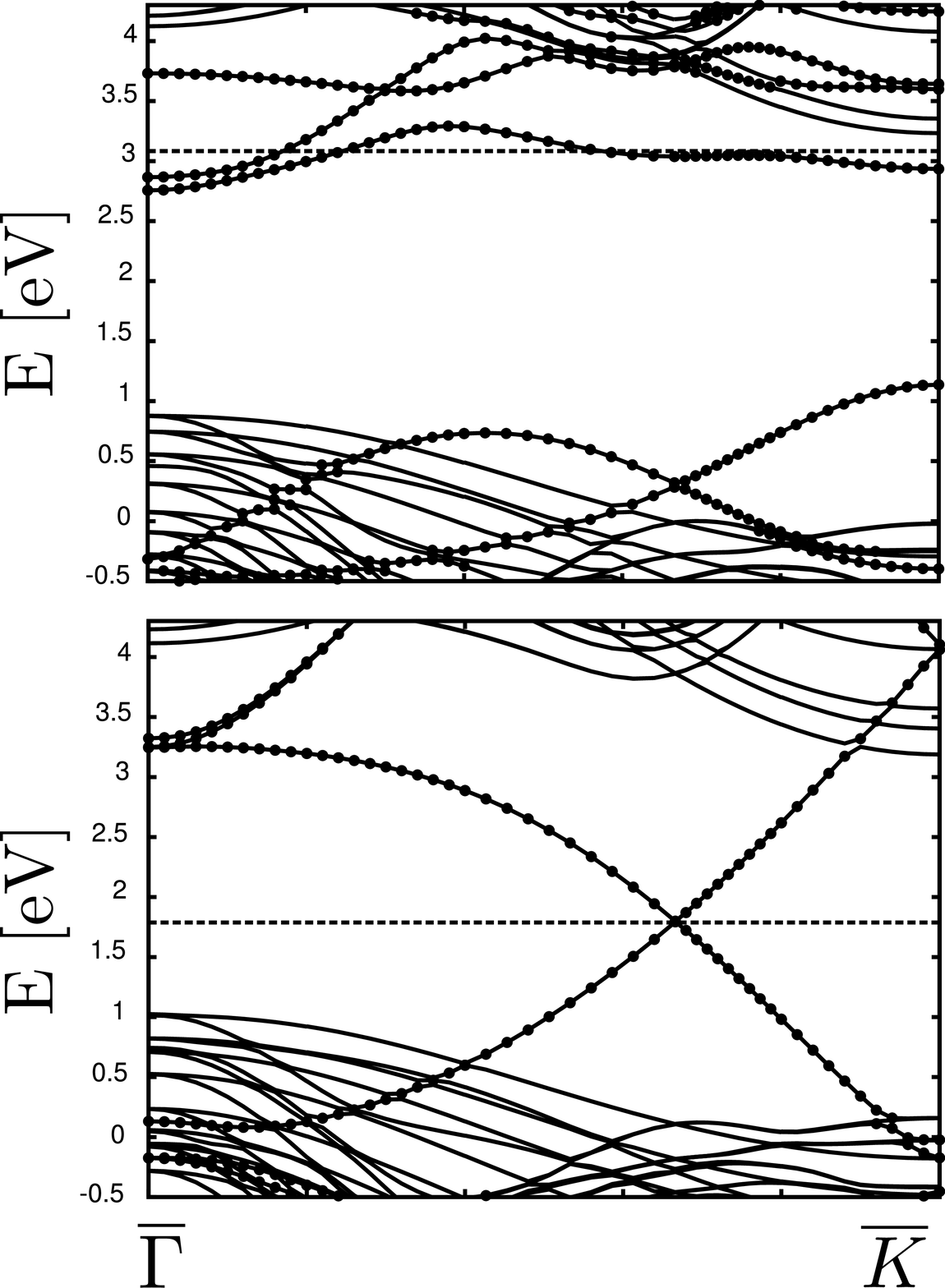}%
\end{center}
\caption{\label{fig:band} Band structure of graphene on SiC($0001$)  
surface with (top) or without (bottom) hydrogen intercalaction atoms.
In the upper case the carbon layer is covalently bound to the surface which removes typical graphene electron properies (Dirac cones are not present in the band structure). In the bottom, the situation completely changed 
due to hydrogen intercalation.Dirac cone is now apparent in the band structure. Also the SiC($0001$) Fermi level intersects Dirac cone. Eigenvalues associated with graphene pz orbitals are marked by circles. 
}
\end{figure}

Resulting band structures of covalently bound graphene on SiC($0001$) surface and quasi free standing graphene
are compared in Fig. \ref{fig:band}. 
Eigenvalues associated with graphene layer are marked with circles.
Band gaps are underestimated which is well known systematic LSDA error.
As described by Mattausch et al. \cite{Mattaush}  the graphene electron spectrum 
is drastically changed by covalent bonding. Graphene Dirac cones are merged into the 
valence band, as can be seen in Fig. \ref{fig:band} - top. 
In the proposed scenarios the hydrogen intercalation breaks the covalent bonding,
leading to recovery of characteristic graphene dispersion cone, see Fig. \ref{fig:band}
- bottom. 

This is in agreement with ARPES measurements performed by Riedl et al. showing
that after hydrogen treatment $\pi$ graphene bands (Dirac cones)  appear.  
Fermi level in this case intersects the Dirac cone.  
Agreement of theoretical (structure, electronic properties) and experimental results (LEED and 
ARPES patterns) suggests that the proposed intercalation scenario is correct.  
This model was previously proposed by Mattausch and Pankratov, but there was not as strong 
(considering situation before and after hydrogen treatment) evidence proving that it is correct. 

Moreover, Riedl et al. pointed out that
the hydrogen intercalated samples are extremely stable in ambient atmosphere,
at least for several months. 
This is also very well explained by our model. Binding energies obtained
from presented calculations demonstrate that configuration with hydrogen atoms bound to the Si surface atoms 
and the graphene layer decoupled is more energetically favorable than coupled graphene-SiC($0001$) and isolated
hydrogen molecule.  That explains high durability and stability of the samples.

The question, left unanswerd by Riedl et al. \cite{Riedl}, is related to the path by which the hydrogen penetrates the graphene layer to arrive at Si surface atoms.
The proposed scenarios include hydrogen migration through graphene lattice or intercalation starting at 
grain boundaries on the surface. 
 
\begin{figure}
\begin{center}
\includegraphics[width=0.7 \textwidth]{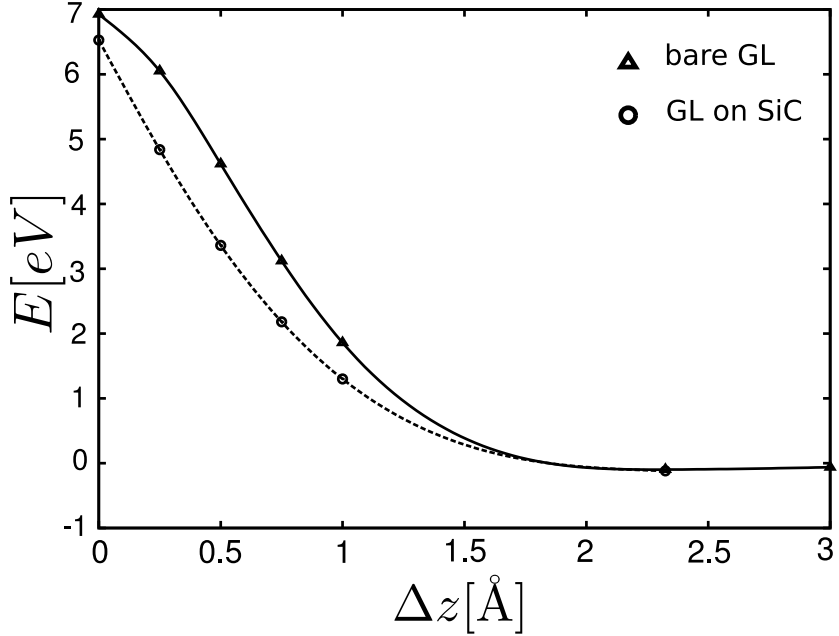}
\end{center}
\caption{\label{fig:h2} Energy adsorption of $H_2$ molecule on 
bare graphene and graphene on SiC($0001$) in function of distance 
between hydrogen and graphene layer. Hydrogen molecule was 
positioned perpendicularly to the surface in center of graphene hexagon and 
above Si-SiC($0001$) dangling bond.
}
\end{figure}

In order to shed more light on the scenario of direct hydrogen penetration of the graphene lattice,  
we have performed another series of calculations. Hydrogen molecule was dragged through 
the graphene layer. Hydrogen was positioned in the center of graphene hexagon perpendicular to
the surface. We have compared two cases of hydrogen adsorbed either on isolated graphene layer or on
graphene on SiC($0001$) surface. In the second case hydrogen atom was placed over 
Si-SiC($0001$) surface dangling bond (see. Fig. \ref{fig:h2} ).
Energy in function of the distance between center of hydrogen molecule and the graphene layer
was presented on Fig. \ref{fig:h2}. In the separate graphene layer case, the energy rapidly increases 
while hydrogen is moving closer with the resulting energy barrier of about 7 eV. In fact SiC surface weakly 
affects the energy profile, reducing the energy barrier to about 6.5 eV. In these two cases, the energy barrier is to high to overcome.

\begin{figure}
\begin{center}
\includegraphics[width=0.7 \textwidth]{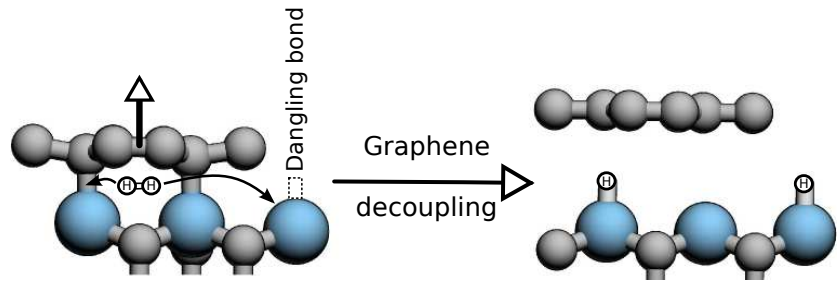}
\end{center}
\caption{\label{fig:dec}
(Color online) Relaxation of hydrogen molecule placed between SiC($0001$) surface and graphene layer \cite{qutemol}. 
Decoupling of graphene takes place in the two steps. In the first step one of hydrogen atoms
is bound to the dangling Si bond. Second hydrogen atom breaks one of SiC-graphene covalent bonds
and is bound to the Si atom. The hydrogen atoms moves perpendicular to the surface, graphene layer 
is decoupled from the surface. }
\end{figure}

Second possibility is that hydrogen intercalation occurs at grain boundaries on the surface. To consider this 
situation we put hydrogen molecule between SiC($0001$) and graphene layer. Relaxation process proceeded in two steps
(see. Fig. \ref{fig:dec}).
In the first hydrogen molecule decomposes and one of hydrogen atom is bound to the dangling bond. 
In the second step free hydrogen atom breaks graphene-SiC($0001$) covalent bond and is bound to Si atom.
Graphene layer is decoupled from the surface as shown in Fig. \ref{fig:hydro} and hydrogen atoms line up 
perpendicularly to the surface. This result shows that it is a possible path of penetrating graphene-SiC($0001$) 
surface system.

In summary, it was shown that the graphene structure is detached from SiC($0001$) surface by intercalated 
hydrogen atoms. Presented results indicate that after hydrogen treatment, the graphene 
properties are recovered (Dirac cone appears in band structure).
The results of the calculations were compared with experimental results \cite{Riedl}. 
The difference of the surface configuration before and after hydrogen treatment is in accordance with
with the experimental (LEED, ARPES) results.  
This compatibility indicates that the used model is correct and 
accurately describes the real structure.
Moreover, we have shown stable graphene-SiC structure penetrated by hydrogen atoms.
We have also considered possible scenarios of hydrogen intercalation - penetration through 
graphene lattice and through grain boundaries. Presented results indicate that moving through
graphene lattice is rather impossible. More probable situation is diffusion through grain boundaries. 
It should be also stressed that graphene layer decoupled from SiC by hydrogen is more energetically favorable
than isolated graphene-SiC system and hydrogen molecule and that intercalation does not destroy 
the graphene structure.
These results explain why such system is extremely stable and that intercalation procedure can be repeated 
\cite{Riedl}. 
Since the intercalation process can recover graphene remarkable electronic properties it can be used to obtain 
homogeneous and large epitaxial graphene layers on SiC($0001$). This is very promising feature 
which can be used in fabrication of nanoelectronic devices.

\section*{Acknowledgements}
The calculations reported in this paper were performed using computing 
facilities of the Interdisciplinary Center for Mathematical and Computational Modeling (ICM) of the University of Warsaw. 
The research published in this paper was supported by Polands Ministry of Science and 
Higher Education grant no. POIG 01.01.02-00-008/08.

\end{document}